# ALGORITHM CERTAINTY ANALYSIS OF SPATIAL DATA FOR TERRAIN MODEL

Vedant A. Garg
Class 12th, DPS RK Puram, New Delhi, India
Email : rkg1969@hotmail.com

Sabir B.Shafi Lone and Swetabh C. Singh
Lovely Professional University/Civil Engineering, Jalandhar, India
IIT Delhi/Civil Engineering, New Delhi, India
Email : sabirshafilone@gmail.com, swetabh2002@gmail.com

*Abstract* --- The terrain survey techniques of photogrammetry, LIDAR, Sonar or seismic studies are subject to limitation of shadow zones. It is not possible to capture the terrain pattern and requires interpolation & extrapolation for conformal mapping of spatial coordinates for generation of terrain model. The discrete data is mapped through function set whose domain returns the analytic test in Riemann map.

The algorithm adopted in analysis for such mapping does not have a certainty score or probability of degree of correctness conforming to the physical landscape of shadow zones.

The aim of the paper is to establish a generator of certainty degree of the mapping along with a continuous terrain model generator. The confirmed mapping of terrain presents a continuous spatial coordinate set which form the boundary of the shadow zone with discrete spatial coordinates. The discrete set is normalized in Gaussian distribution through a Poisson distribution transition. The continuous data set is represented by Laurentian series in which the function will be analytic and can be mapped to Riemann surface with singularities within the annulus and outside the annulus of approximate space sub set (Euclidean space). The singularities will be discarded through Picard's Theorem and analytic test at poles with Cauchy's residual theorem. The resulting set of spatial coordinates will restructure within Riemann number sphere which will be mapped on the plane as stereographic projection.

The Gaussian distribution which forms the basis of analysis will provide with the tool for generating the probability of certainty of every terrain model idealised to conform to the physical landscape.

Remote sensing methods are based upon transmitting of transverse pulses to the terrain & measuring the return time of the associated backscattered waves. The slant measurements & altitude is used to derive the three dimensional point clouds which is the universal set of geo reference points.

The ground water, submerged surfaces, forest, convoluted surfaces, discontinuities etc. are captured by the wave front. The projection of the features is dependent upon the processing algorithm. These algorithm need to be developed for accounting the scattering, absorption effects of the wave front to generate a real three dimensional model with least approximation.

The sources of error include interpolation from discrete ground to continuous model. The interpolation errors can be deciding factor as none of the established algorithm capture the negative curvatures & corner projections with close to reality or depicting real ground model. It has been found that the algorithm eliminate points at corner between near vertical faces or negative curvatures as are in hilly terrain of steep mountainous region(Himalayan Region) with deep valleys & gorges, steep slope of the mountains, landslides, glaciers, roads etc.

Some of the algorithm that has been widely used is:-

1. Despite algorithm
2. Multiple return block algorithm
3. Block minimum algorithm
4. Iterative linear prediction

1- **Despite Algorithm**
   Despite algorithm has been widely successful and the return wave front is examined by representing the terrain in Triangulated irregular network but suffers from the problem of neglecting the points of strong curvatures or convexities but still it is an advantage over iterative interpolation algorithm which smoothens out all the curves.

2- **Multiple Return Block Algorithm**
   Multiple block return algorithm has been deployed successfully for bare ground model but does not return the terrain model for ground surface beneath the canopy, curvatures, corners etc.

3- **Block Minimum Algorithm**
   Block minimum algorithm is based upon the ground points being the lowest in the neighbourhood. The derived surface will have biased towards low or steep slopes than in reality.

4- **Iterative Linear Prediction**
   Iterative linear prediction attempts at modelling the ground surface by linear interpolation & iterate the data to fix the points in space. It uses the degree of smoothness of ground to build a ground surface model.

The focus of the paper is to discuss the limitations of interpolation methods in algorithm. Most advances in interpolation methods have been through the finite difference method and algebraic polynomial interpolation for numerical solution of Poisson's Equation over any domain of investigation. FDM uses topologically square network of lines for discretised data & thus the difficulty in handling complex geometries in multiple dimensions. The kind of interpolation methods used is:-

1- **Inverse distance weighted**
   The interpolator assumes that each input point has a local influence that diminishes with distance. It weights the point closer to processing cell greater than those further away. Use of this method assumes the variable being mapped decreases in influence with distance from its sampled location.

2- Natural neighbour inverse distance weighted is used for interpolation & extrapolation & works well for clustered scatter points. The local co-ordinates define the amount of influence any scatter point will have on output cells. This is a geometric interpretation technique that uses natural neighbourhood regions generated around each point in the data set. The algorithm creates a triangulation of the input points & selects the closest nodes that form a convex space around the interpolation point then weights the value by proportionate area. This method is most appropriate where sample data points are distributed with uneven density.

3- SPLINE estimates values using mathematical function that minimizes overall surface curvature smothering the surface to idealised equations. This is best suited for gently varying surfaces.

4- KRIGING -: It is a geostatistical interpolation technique that consists both the distance & the degree of variation between known data points when estimating value in unknown area.
   It generates an estimated surface from a scattered set of points with Z-values.
   The Kriging formula

$$Z(S_o) = \sum_{i=1}^{N} \lambda_i\, Z(S_i)$$

$Z(S_i$ : the measured value at $i^{th}$ location

$\lambda_i$ : an unknown weight for the measured value at the $i^{th}$ location

$S_o$ : Prediction Locator

N : Number of the measured value

5- TREND is a statistical method that finds the surface that fits the sample points using a least square regression fit. It fits one polynomial equation to the entire surface. This results in a surface that minimizes surface variance in relation to the input values.

As has been mentioned regarding extensive usage FDM with algebraic polynomial interpolation or interpolation polynomial in the LaGrange form

$$L(x) = \sum_{j=0}^{K} y_i \, l_j(x)$$

Or the use of Newton polynomial

$$N(x) = \sum_{j=o}^{K} a_j n_j(x)$$

Also referable as Newton's divided difference interpolation polynomial where the FDM fits perfectly. These algorithms suffer from the inherent drawback of being uncertain at discontinuities. As they all operate at collecting discrete points & generating a continuum making their behaviour at discontinuities, convexities boundaries completely unrealistic or divergent.

The basic motivation for working with polynomials is the convenience with which it can be readily differentiated or integrated for any continuous $f(x)$ on an interval, j:a≤x≤b an error bound $B≥0$ there is a polynomial $Pn(x)$ of high degree n such that $|f(x)-Pn(x)|<B$ for all $x$ on $J$. This is the famous Weierstrass approximation theorem. The error estimate for polynomial of degree n, it must coincide with the $P_n$ because $n^{th}$ data $(x_o,f_o)…………..(x_n, f_n)$ determine a polynomial uniquely so that the error is zero. Now the special $f$ has its nth derivative identically zero. This makes it plausible that for a general $f$ its (nth) derivative should measure the error.

$$\in n(x) = f(x) - Pn(x)$$

$$= (x - x_o)(x - x1) … … … … (x - x_n) \frac{f^{n+1}}{(n+1)!}(t)$$

This $|\in n(x)|$ is 0 at the nodes and small near them, because of continuity. The term $(x-xo)……(x-xn)$ is larger for x away from the nodes. This makes the extrapolation & interpolation on boundaries risky as the deviations or errors recorded will be very high. Similarly in Newton's forward difference interpolation formula.

$$f(x) = Pn(x) = \sum_{s=0}^{n} \frac{r}{s} \Delta^2 fo$$

$$= fo + r\Delta fo + \frac{r(r-1)\Delta^2 fo}{2!}$$
$$+ \frac{r(r-1)…………(r-n)\Delta^2 fo}{n!}$$

Where the coefficients are binomial expressions.

Error.
$$\in n(x) = f(x) - Pn(x)$$
$$= \frac{h^{n+1}}{(n+1)!} r(r-1)……(r-n) f^{n+1}(t)$$

Show that the error will assume high proportion for intervals beyond certain value & observations beyond certain value.

$$\in n(\ddot{x}) = 0 = \frac{(n+1)h^n h r(r-1)….(r-n) f^{n+1}(t)}{(n+1)!}$$

For limiting value $\sqcup n \to$

Which clearly indicates that for greater observations & negative gradation (h) the error is maximised which generally is the case at boundaries.

This is the reason that all the algorithm with real function, polynomial models for interpolation will suffer with inherent defect of error maximisation on divergent boundaries or discontinuities.

## SIGNIFICANCE OF THE USE OF COMPLEX FUNCTION THEORY

Real numbers form a topological space and a complete metric space. Continuous real-valued functions are important as they are deployed in topological spaces and of metric spaces also they are used in complex analytical functions satisfying Cauchy Riemann theorem. The extreme value theorem states that for any real continuous function on a compact space its global maximum and minimum exist. The concept of metric space itself is defined with a real-valued function of two variables, the metric, which is continuous. The space of continuous functions on a compact Hausdorff space has a particular importance. Convergent sequences also can be considered as real-valued continuous functions on a special topological space.

An example will explain the stated position.

For any variable point in the neighbourhood of the topological manifold the points can be defined algebraically and following condition will be derived.

$$2 < |2 + iy| < 3$$
$$0 < y < \pm\sqrt{5}$$

The terrain function can be generated with statistical model by any generator which in this case has been assumed considering the continuum. The error of assumption will be common for polynomial interpolator as well as complex interpolator as the assumption is consistent in both the models.

$f(x) = x^2+x+1$ $x(2,3)$ close interval, $z(2,3)$ close interval

$$Y = \frac{14 \pm i\sqrt{54}}{4}$$

$f(x) = x^2+x+c$ will return values for homeomorphism in landscape.

$F(Xo)$ the map will not reflect the terrain map $T(Xo)$

The error function will be computed upon terrain model $E=f(Xo)-T(Xo)$

X can assume non real roots. For all $x(0,-2)$ $f(x)$ will assume a smooth map where $f(x)=0.75$ for $x=-0.5$, for $T(x)=-10$

$E(Error)=-3/4+(-10)=-10.75$

For $z=0.5+i3.4$, $f(z)=11.93$

As is evident with complex interpolator the error has been reduced from 900% to 16%.

The real valued function are not flexible to use as the domain set & topological space are not continuous in neighbourhood of variables values assigned a polynomial function or quadratic function $f(x) = x^2+x+c$ for the real function $X$ can be $X \in (-\infty,\infty)$ $f(x) \in (-\infty,\infty)$ i.e. the domain and range will be $\subset$ of R.

For any real valued function in a domain space $a \le x \le b$ will have a range of the function $A \le f(x) \le B$ with restriction on $x$ & Range of $f(x)$. The restrictions in terms of the applicability of the real valued function for topological discontinuities where the real value function will be non-analytical.

The scope of this discussion is to establish that the flexibility offered by complex functions is arriving at or minimising the error in topological map is much more than being offered by the real analytical function. The applicability of complex functions in the given domain of convex manifold or geodesic convexity will interpolate to the precision of topography & thus we can introduce a complex interpolation function.

$f : R \to R$ is the real valued analytic function for the interpolation such that $f : X \to Y$ $X \subset R$ & $Y \subset R$ for any convex topology or discontinuity there exists. T: $Z \to W$ $Z \subset C$ & $W \subset C$ $\Delta f: Z \to W$ such that $\lim \Delta f \to 0$ in all conformal mapping $Z \to W$

## ENERGY FUNCTION & POYNTING VECTOR

The electromagnetic waves, lasers, Radio waves emitting from the source for imaging of the terrain has parameter with control. Generally spherical wave fronts are generated and the reflected & refracted values are studied to map the terrain.

The energy function as the rate of energy transport per unit area is the poynting vector. The absorption & reflection on time variation will provide the differential plot of the ground surface which shall be closed with the interpolator. For spherical wave fronts the intensity function which is the poynting vector time averaged as

$$I = \frac{E_{rms}^2}{c\mu_o}$$

$E_{rms}$ is the electric field root mean square

However as the intensity varies with distance from the real source of electromagnetic radiation. Assuming the source of radiation to be isotropic the spherical wave fronts are more suited for convergence. The energy of the wave front is conserved.

So for a source of power P

$$I = \frac{P_s}{4\pi r^2}$$

Which clearly implies that

$$E_{rms} = \sqrt{\frac{P_s c \mu_o}{4\pi r^2}} = \frac{k}{r}$$

The range of radiation pressure for laser pulse from $\frac{1}{c} \leq pr \leq \frac{27}{c}$ provides the judgement for the absorbed and reflected or refracted waves for modelling the terrain.

## INTERPOLATOR & CONVERGENCE OF THE FIELD

The field function *f(z)* with singularity cannot be defined with Taylor series & requires Laurent Series which helps classify the singularities & key to integration of residue for geometric convergence of the terrain model.

$$f(z) = \sum_{n=0}^{\infty} a_n (z - z_o)^n$$

$$+ \sum_{n=1}^{\infty} \frac{bn}{(z - z_o)^n}$$

$$a_n = \frac{1}{2\pi i} \oint_c \frac{f(z^*) dz^*}{(z^* - z_o)^{n+1}}$$

$$b_n = \frac{1}{2\pi i} \oint_c (z^* - z_o)^{n-1} f(z^*) dz^*$$

The field equation can be written as:

$$a_n = \frac{1}{2\pi i} \oint_c \frac{k/r \, (dr)}{(r - r_o)^{n+1}}$$

$$b_n = \frac{1}{2\pi i} \oint_c (r - r_o)^{n-1} \frac{k}{r} dr$$

The function converges into

$$\sum_{i=0}^{\infty} \frac{k \ln r_i}{r_i - r_o} = f(z)$$

$r_o$ will be the average radius of the wave front or the distance from the source to terrain.

$$\frac{k \ln r_i}{r_i - r_o} = E_i = \sqrt{\frac{PC\mu_o}{4\pi r_i^2}}$$

Thus the equation can be used for interpolation of $r_i$ or the interpolation function in $r_i$.

$$\frac{\ln r_i}{(1 - \frac{r_o}{r_i})} = C$$

The logarithmic functions have an inverse function (1/r) differential which for higher values of the radius of wave fronts will converge the error. The truncated solution forms the closed approximation of the solution and can be evaluated at any point within the region except at the pole. Instability of finite difference method does not affect the solution.

## CONCLUSION

The convergence in logarithmic series function is reachable with the observation frequency expected in the field investigation. The singularities have not been undermined & have been taken care of by residue integration theorem. The elegance in the solution has been reached on the basis of analytic function of complex variables which is the universal set of all numbers & the solution is in Euclidean-space making it globally acceptable & usable for computation as it has been given a general form. The solution is based upon the wave form (spherical) in present case which is in contrary to Lagragian & Newtonian equations which are based upon polynomial functions only which is adopted irrespective of the wave propagation equation & hence the effort for convergence is always high & inaccurate due to inconsistency of parameters. The importance of Laurent series of a given analytic function *f(z)* in its annulus of convergence is unique even if *f(z)* may have different Laurent series in two annuli with same centre thus the uniqueness in solution with consistency & easy convergence makes it a better interpolation function.

The similar convergence can be drawn using the cylindrical, planar wave front and develop the interpolation for profiling the terrain.

**Vedant A. Garg** is an Indian born 12th grader currently studying in Delhi Public School R.K. Puram in New Delhi. Born at Giridih, Jharkhand in India on 5 April 2002, Vedant has selected the science stream and studies Physics, Math, Chemistry and Computer Science in school.

He has worked with several organisations in a short time. An avid environmentalist, he worked as the HEAD OF BUSINESS OPERATIONS in Delhi to create a network of Electric Vehicle charging stations and made them accessible through an app available on both the App Store and Play Store. As the IT AND SOCIAL MEDIA HEAD of Save Child Beggar, a youth lead organisation to educate the less fortunate, he has helped launch it to become one of the largest social outreach organisations in the country. Vedant also worked with Majime as a RESEARCHER and developed a mathematical model to show the effects of a reputation model on the gig economy. The company has finished work on a blockchain powered platform and hopes to get it patented very soon. His interest lies in mathematics and computation and he hopes to use it to solve real world issues. Mr. Garg is an active member of his school community and is part of several societies. He has won several aerospace competitions organised by NASA. He is also an NTSE Scholar, an exam with 0.01% qualification rate in India and also a holder of the prestigious KVPY Fellowship given by the Indian Institute of Science (IISc).

**Sabir B. Shafi Lone** born in city of Baramulla in the state of J&K, India on 29th of January, 1992. The author completed his school education in Baramulla city only. The author completed his bachelor of Technology from Lovely professional University in the year 2015 and Post graduate programme from NICMAR (SODE), in year 2018 in field of Construction Management.

The author has worked extensively in the Educational and infrastructure industry for more then 03 years on complex projects like Construction of bridges and Highways and teaching subjects like Project management, RCC, Earthquake engineering and many more working with organisations like Lovely Professional University, NCSL-MTDCL(JV). The engagement in teaching career was related to teaching of different subjects and research projects on same while engagements in company were limited to planning and designing roles.

The Author has published in year 2016 in Indian journal of Science and Technology on Comparative study of Nano Silica Induced Mortar, Concrete and Self-Compacting Concrete – A Review.

**Swetabh C. Singh** born in city of Gorakhpur in the state of Uttar Pradesh, India on 29th of October, 1985. The author completed his school education traveling across India in State of Uttar Pradesh, West Bengal, Sikkim, Gujarat, Bihar, Delhi as the father of author was in All India service of Government of India and had to be transferred frequently. The author completed his bachelor in Technology from National Institute of Technology, Allahabad in the year 2006 and Master in Technology from Indian institute of Technology, Delhi in year 2014 in field of Rock Engineering and underground structures with special interest in true triaxial stress analysis and plastic strength criterion for Rock Mass and mobilisation of ultimate capacity of Rock Mass for tunneling and caverns.

The author has worked extensively in the infrastructure industry for more then 12 years on complex projects like Delhi International Airport , Paradip Refinery project, LNG petronet Dahej for jetty development, Chenab River bridge, Rohtang Tunnel project, Sylkiyara barkot bend tunnel and many more, working with organisations like Larsen & Toubro LTD, Fluor Daniel, Indian oil tanking LTD, IIT Delhi, and as an Independent consultant. The engagement in most of the projects involved in capacity of Design Engineer or geotechnical Engineer.

The Author has published in year 2007-08 in 58th IAC conference in smart material & adaptive structures symposium on Intelligent defect Genesis. The current research interests are triaxial strength criterion for geomaterials, mathematical modelling and structural geology.